\documentclass[a4paper,11pt]{article}
\pdfoutput=1 
\usepackage{color}
\usepackage{jcappub} 

\usepackage[T1]{fontenc} 
\usepackage{threeparttable}
\usepackage[labelformat=empty]{subfig}
\usepackage{booktabs}
\usepackage{multirow}
\usepackage{placeins}
\usepackage[utf8]{inputenc}
\usepackage{indentfirst}
\usepackage[utf8]{inputenc}

\title{\boldmath Observing EeV neutrinos through the Earth: GZK and the anomalous ANITA events}

\author[a]{Ibrahim Safa,}
\author[a]{Alex Pizzuto,}
\author[b]{Carlos A. Arg{\"u}elles,}
\author[a]{Francis Halzen,}
\author[a]{Raamis Hussain,}
\author[a]{Ali Kheirandish,}
\author[a]{Justin Vandenbroucke}

\affiliation[a]{Dept. of Physics and Wisconsin IceCube Particle Astrophysics Center, University of Wisconsin, Madison, WI 53706, USA }
\affiliation[b]{Dept. of Physics, Massachusetts Institute of Technology, Cambridge, MA 02139, USA }

\emailAdd{isafa@wisc.edu}
\emailAdd{pizzuto@wisc.edu}
\emailAdd{caad@mit.edu}
\emailAdd{halzen@icecube.wisc.edu}
\emailAdd{rhussain@icecube.wisc.edu}
\emailAdd{akheirandish@icecube.wisc.edu}
\emailAdd{justin.vandenbroucke@wisc.edu}

\abstract{Tau neutrinos are unique cosmic messengers, especially at extreme energies. When they undergo a charged-current interaction, the short lifetime of the produced tau gives rise to secondary tau neutrinos that carry a significant fraction of the primary neutrino energy. This effect, known as tau neutrino regeneration, has not been applied to its full potential in current generation neutrino experiments. In this work, we present an updated calculation of tau neutrino regeneration, and explore its implications for two scenarios: the recent anomalous ANITA events and the cosmogenic neutrino flux. For the former, we investigate the idea of localized emission and find that the maximum secondary neutrino flux allowed by IceCube measurements implies a primary flux that is incompatible with the ANITA observation, regardless of the assumed source energy spectrum. For the latter, we study the prospect of detecting the cosmogenic neutrino flux of regenerated PeV neutrinos with current and next generation neutrino detectors.}

\notoc 
\begin{document} 
\maketitle
\flushbottom

\section{Introduction\label{sec:intro}}

Neutrinos provide the opportunity to probe the most cataclysmic and energetic processes in the Universe. As they are immune to magnetic fields, and their interactions with matter are extremely feeble, high-energy neutrinos may reach us unscathed from the edge of the Universe. However, as was pointed out since the neutrino's inception \cite{Pauli:1930pc}, the smallness of the neutrino cross section is a double-edged sword, as the remarkable ability of neutrinos to escape dense astrophysical environments goes hand in hand with the ability to pass through detectors~\cite{Gaisser:1994yf}. The neutrino detection problem becomes even more challenging for rare neutrino production processes. The two most elusive predicted neutrino fluxes are the cosmic neutrino background (C$\nu$B) and the cosmogenic flux. The former is the largest flux of naturally produced neutrinos. Unfortunately, it peaks at meV energies, where its cross section has made it undetectable as of yet. The latter is a guaranteed but yet to be detected flux of extremely-high-energy (EHE) neutrinos produced in weak decays of particles from the interactions of ultra-high-energy cosmic rays (UHECR) with the cosmic microwave background (CMB)~\cite{Greisen:1966jv, Zatsepin:1966jv, Beresinsky:1969qj}. The cross section around these energies reduces the interaction length of neutrinos to $\mathcal{O}(100)$ km in rock, but the flux is $\sim$50 orders of magnitude smaller than the C$\nu$B, making it equally elusive.
Soon after the prediction of the cosmogenic neutrino flux, its level made it evident that cubic kilometer detectors are required to observe this flux at high energies~\cite{Roberts:1992re, Halzen:2008zz}. Later estimates for observing potential cosmic
accelerators such as Galactic supernova remnants and gamma-ray bursts
pointed to a similar requirement~\cite{Gaisser1995,Learned:2000sw,Halzen:2002pg}. 

The discovery of astrophysical neutrinos by IceCube marked the beginning of high-energy neutrino astro-particle physics \cite{Aartsen:2013jdh}. This initial observation was followed by the detection of an excess of a high-energy astrophysical muon-neutrino flux component above the atmospheric background in the northern sky \cite{Aartsen:2015rwa}. These initial measurements have been confirmed recently with 9.5 years of northern sky muon-neutrino data \cite{Stettner:2019tok} and 7.5 years of all-sky starting event data \cite{Schneider:2019ayi}. 
The astrophysical flux observed by IceCube saturates the theoretical flux expectations \cite{Waxman:1998yy} and is predominantly extragalactic \cite{Albert:2018vxw}. Intriguingly, as well, the total energy density in high-energy neutrinos is similar to the energy density of the UHECRs which might hint at their common origin. This observed flux, however, is not the cosmogenic neutrino flux, and the predominant sources are yet to be identified. 

In the search for cosmogenic neutrinos, IceCube selects the highest energy depositions corresponding to $\sim$EeV events. The main backgrounds in this region are the astrophysical component and muons produced in cosmic-ray showers. To reject the latter, a zenith-angle dependent cut on the deposited energy is applied resulting in the largest sensitivity near the horizon. Additionally, the Earth shields the detector from a large fraction of the primary cosmogenic flux in the northern sky due to the growing neutrino cross section ~\cite{Gandhi:1998ri,Connolly:2011vc,Vincent:2017svp}. Therefore the search is eventually limited to a region near the horizon; a sliver of the full sky.
Similarly, neutrino detectors sensitive to higher energies compared to IceCube
have typically limited searches 
to Earth-skimming or downgoing trajectories, where the column depth is optimal for detecting EHE neutrinos after a single interaction~\cite{Kotera:2010yn}. Experiments such as ANITA, ARA, ARIANNA, and the Pierre Auger Observatory have set limits on the cosmogenic flux taking advantage of the Earth-skimming technique~\cite{Abbasi:2010ak, Barwick:2014pca, Aab:2015kma, Aartsen:2016ngq, Allison:2015eky,Allison:2018cxu,Aartsen:2018vtx,Aab:2019auo}.

Proposed experiments such as RNO, GRAND, CHANT, POEMMA, and IceCube Gen-2 \cite{Aguilar:2019jay, Fang:2017mhl, Neronov:2016zou, Krizmanic:2019hiq, Aartsen:2014njl} rely on the Earth-skimming technique for detection of EeV neutrinos. However, tau neutrinos offer a unique opportunity to detect those neutrinos which prematurely interact in the Earth prior to reaching the detector. When an incident tau neutrino undergoes a charged-current (CC) interaction, the subsequent decay of the produced tau will yield another tau neutrino at a lower energy~\cite{Learned:1994wg, Halzen:1998be,Beacom:2001xn, Dutta:2002zc,Bugaev:2003sw,Dutta:2005yt,Reno:2019jtr}. Although this process is not unique to the tau channel, the energy distribution of the secondary neutrinos peaks at much higher energies for tau neutrinos than for muon- or electron- neutrinos. This effect has been dubbed "tau neutrino regeneration," and will be discussed further in Sec.~\ref{sec:leptons}. 

In this paper, we take advantage of tau neutrino regeneration to study EHE neutrino fluxes by looking at their resultant secondaries. Numerous calculations have been performed with different approximations to solve the tau neutrino transport problem \cite{Bugaev:2003sw, Dutta:2000jv, Beacom:2001xn, Dutta:2002zc, Yoshida:2003js, Bottai:2002nn, Giesel:2003hj, Reno:2019jtr}. In our treatment of tau neutrino propagation, we include tau energy losses and show that signatures of Earth-traversing neutrinos provide an opportunity to infer a neutrino flux at EeV energies through its secondaries. Using this technique, we extend the parameter space to a previously neglected region below the horizon and discuss the prospects of detecting cascaded neutrino fluxes, or Earth-traversing EHE neutrinos. For this purpose, we develop a Monte Carlo software package, \texttt{TauRunner} and describe it in Sec.~\ref{sec:tau_runner}. 

In Sec.~\ref{sec:secondary_flux} we show that Earth-traversing EeV neutrinos emerge at $\mathcal{O}$(PeV) energies, a region where IceCube has already performed measurements. We further highlight the connection between EeV and PeV regions by investigating the recent anomalous EeV events reported by ANITA. 
ANITA is a radio-balloon experiment that flies over the Antarctic ice in search of the cosmogenic neutrino flux. During the third flight of ANITA in 2014, an event (AAE141220) was detected which appeared to be an upgoing tau shower initiated by a tau neutrino interaction in the ice. The reconstructed direction, however, implies a column depth through the Earth corresponding to $\sim$20 interaction lengths for an EeV neutrino. The implied survival probability coupled with an isotropic emission assumption requires a flux that is in tension with cosmogenic neutrino limits \cite{Romero-Wolf:2018zxt, Fox:2018syq, Chipman:2019vjm}. But, discrete source emission can evade these bounds. In section \ref{sec:ANITA}, we prove for the first time that any localized emission which would result in AAE141220 is in severe tension with IceCube measurements at PeV energies, closing the last loophole in the neutrino interpretation of the ANITA events.

In Sec.~\ref{sec:GZK}, we propagate a cosmogenic neutrino flux model \cite{Ahlers:2010fw} through the Earth. We find that the rate of Earth-traversing tau neutrinos expected at IceCube is twice the rate of Earth-skimming events, with a well understood and unique energy and zenith distribution. These handles will allow separation from atmospheric and astrophysical backgrounds in future dedicated analyses. Finally, we discuss the current strengths and limitations of this approach as well as future prospects for IceCube Gen-2 in Sec.~\ref{sec:conclusion}.

\section{Leptons through the Earth\label{sec:leptons}}

The propagation of a flux of neutrinos through a medium can be described by the following cascade equation~\cite{GonzalezGarcia:2005xw}
\begin{equation}
    \frac{d \varphi(E, x)}{d x}=-\sigma(E) \varphi(E, x)+\int_{E}^{\infty} d \tilde{E} ~  f(\tilde{E}, E) \varphi(\tilde{E}, x),
\end{equation}
where $E$ is the neutrino energy, $x$ is the target column density, $\sigma(E)$ the total neutrino cross section per target nucleon, $f(\tilde{E}, E)$ is a function that encodes the migration from larger to smaller neutrino energies, and $\varphi(E, x)$ is the neutrino spectrum.
The first term on the right hand side accounts for the loss of flux at energy $E$ due to charged-current (CC) and neutral-current (NC) interactions, whereas the second term is the added contribution from neutrinos at larger energy, $\tilde{E}$, to $E$ through NC interactions of $\nu_{e, \mu, \tau}$ and CC interactions in the $\nu_{\tau}$ channel.
In this work, the secondaries produced in CC interactions of other flavors are neglected due to the fact that the electrons and muons lose  energy rapidly.
On the other hand, taus produced in CC tau neutrino interactions have a much higher probability of decaying yielding high-energy neutrinos.
This is due to the fact that weak decays scale as $m^5$ and that the tau mass is significantly larger than that of the muon, allowing for more decay modes, which results in a ratio of lifetimes between muons and taus of approximately $10^7$.
While the lifetimes are drastically different, the energy losses above $\sim$1 PeV, where stochastic losses are dominant, are only a factor of 10 smaller for taus than for muons.
These two facts set the critical energy in ice -- the energy at which the decay and interaction lengths are equal -- to be approximately ${\sim} 10^9$ GeV for taus, while for muons it is ${\sim}10$ GeV~\cite{Koehne:2013gpa}. 
This implies that tau energy losses can be safely neglected below 10 PeV and the on-spot instantaneous tau decay approximation is a good one, see {\it e.g.}~\cite{Vincent:2017svp}.
However, in this work we consider neutrino propagation at EeV energies and higher, where this approximation no longer holds and careful treatment of tau energy losses is required; see~\cite{Alvarez-Muniz:2017mpk, Reno:2019jtr} for recent implementations and discussions. 

\subsection{Lepton behavior at extremely-high-energies\label{sec:leptons_ehe}}

Measurements of neutrino cross sections have been performed from sub-GeV up to a few PeV energies~\cite{Patrignani:2016xqp}.
This includes a multitude of results using human-made neutrinos in accelerator~\cite{AguilarArevalo:2010zc,Tzanov:2005kr} and reactor~\cite{Vogel:1999zy,Kurylov:2002vj} experiments, as well as natural sources such as solar~\cite{Agostini:2018uly}, atmospheric~\cite{Li:2017dbe}, and astrophysical neutrinos~\cite{Aartsen:2017kpd,Yuan:2019wil}; for recent reviews see~\cite{Formaggio:2013kya,Katori:2016yel}.
In the future, measurements of high-energy neutrinos from collider experiments will be available in the TeV range~\cite{Feng:2017uoz,Abreu:2019yak}.

\begin{figure}[htb!]
    \centering
    \includegraphics[width=\linewidth]{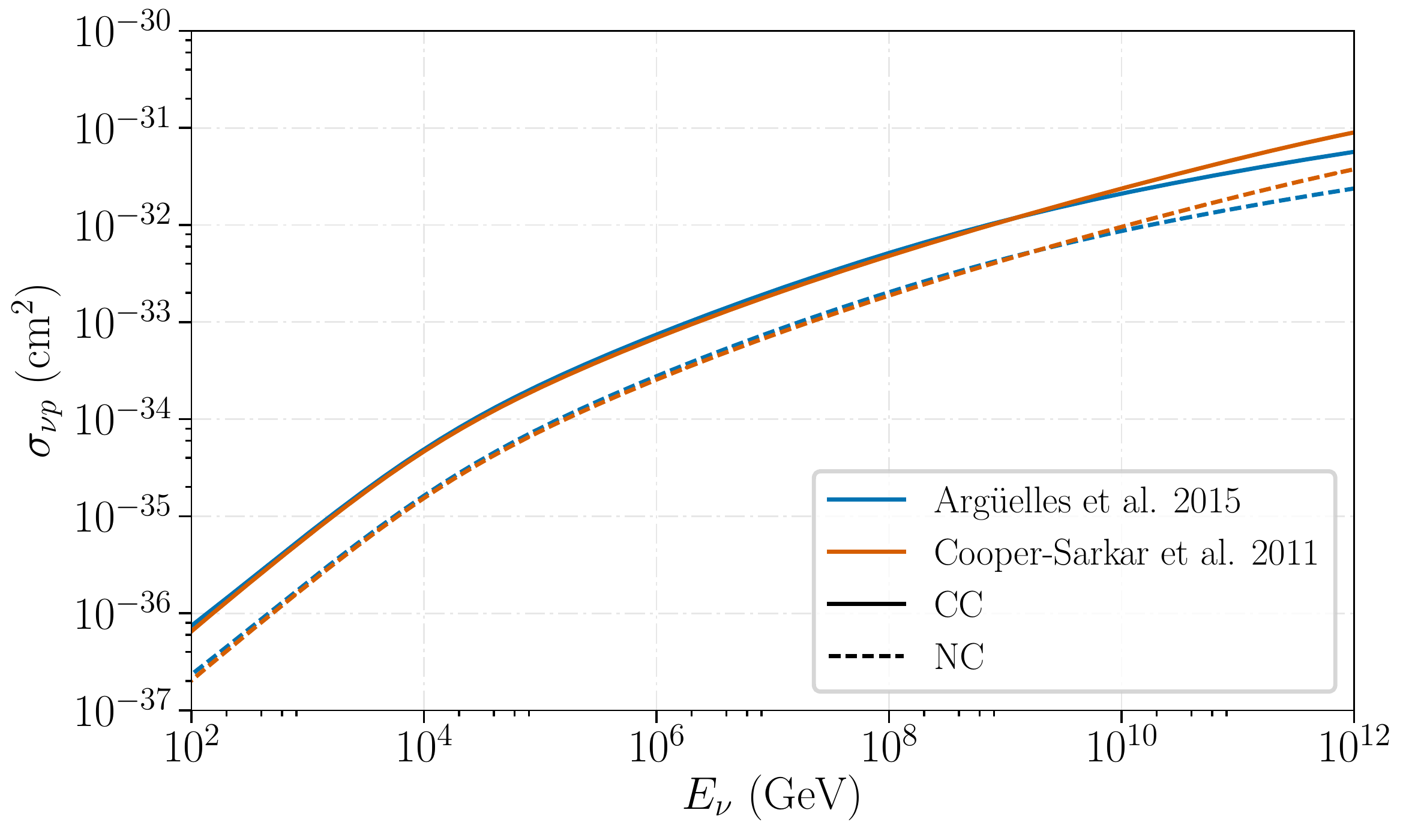}
    \caption{The neutrino-proton cross section as a function of the neutrino energy. Solid (dashed) lines correspond to charged-current (neutral-current) cross sections. Blue lines~\cite{Arguelles:2015wba} correspond to the model used for this work. Orange lines~\cite{CooperSarkar:2011pa} are implemented in the software as well and can be chosen by the user.}
    \label{fig:nucross}
\end{figure}

Unfortunately, these measurements stop short of the region of interest for this work and predictions of the very-high-energy neutrino cross sections disagree at the highest energies; see Fig.~\ref{fig:nucross}.
The main issue driving these uncertainties is that the nucleon structure functions cannot be derived from first principles, which causes us to instead rely on empirical measurements.
Perturbative QCD calculations of the high-energy neutrino cross section are in good agreement with each other when physical consistency requirements are imposed on the PDFs used~\cite{Gandhi:1998ri,Connolly:2011vc,CooperSarkar:2011pa,Bertone:2018dse}, however they grow at a rate $\left(E_\nu^{0.3}\right)$ that will eventually violate the Froissart bound~\cite{Froissart:1961ux, Block:2011vz, Block:2004ek}. This unphysical growth is due to extrapolation of the PDFs to unmeasured phase space. A phenomenological approach~\cite{Block:2000gy, Block:2014kza} to address this issue relies on a $\ln^2(s)$ extrapolation of low-energy measurements using a dipole model of the nucleon. Calculations using this approach were shown to be in good agreement with the total proton-proton cross section measurements from Auger~\cite{Collaboration:2012wt} and TOTEM at LHC~\cite{Antchev:2013gaa} data. In this work, we use the dipole model calculation given in~\cite{Arguelles:2015wba} as our model for neutrino-nucleon interactions; this results in structure functions compatible with~\cite{Block:2014kza}.

\begin{figure}[!t]
   \centering
   \subfloat[][]{\includegraphics[width=.46\textwidth, trim={1.5cm 1.75cm 2cm 2cm}]{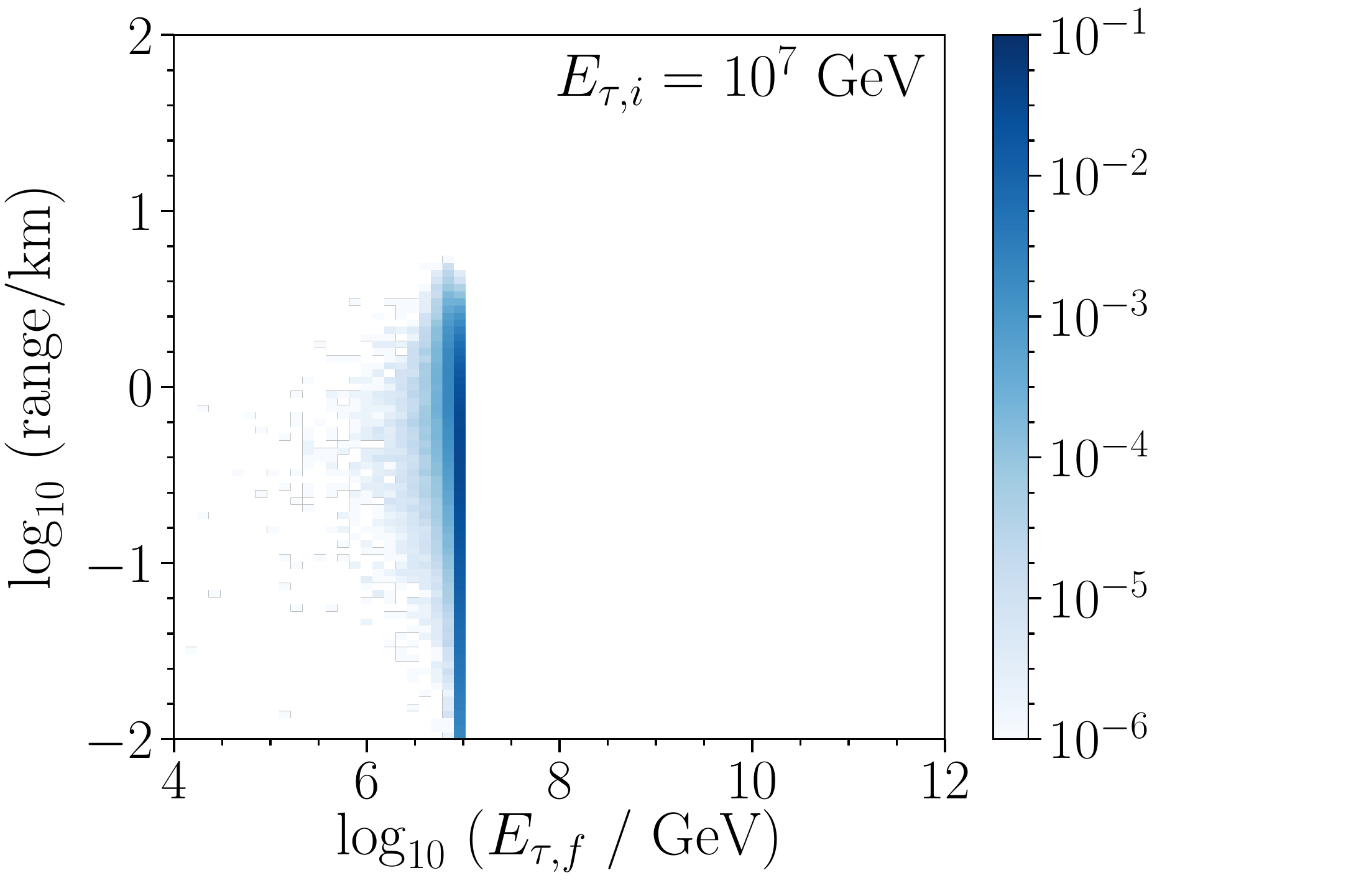}}\quad
   \subfloat[][]{\includegraphics[width=.46\textwidth, trim={1.5cm 1.75cm 2cm 2cm}]{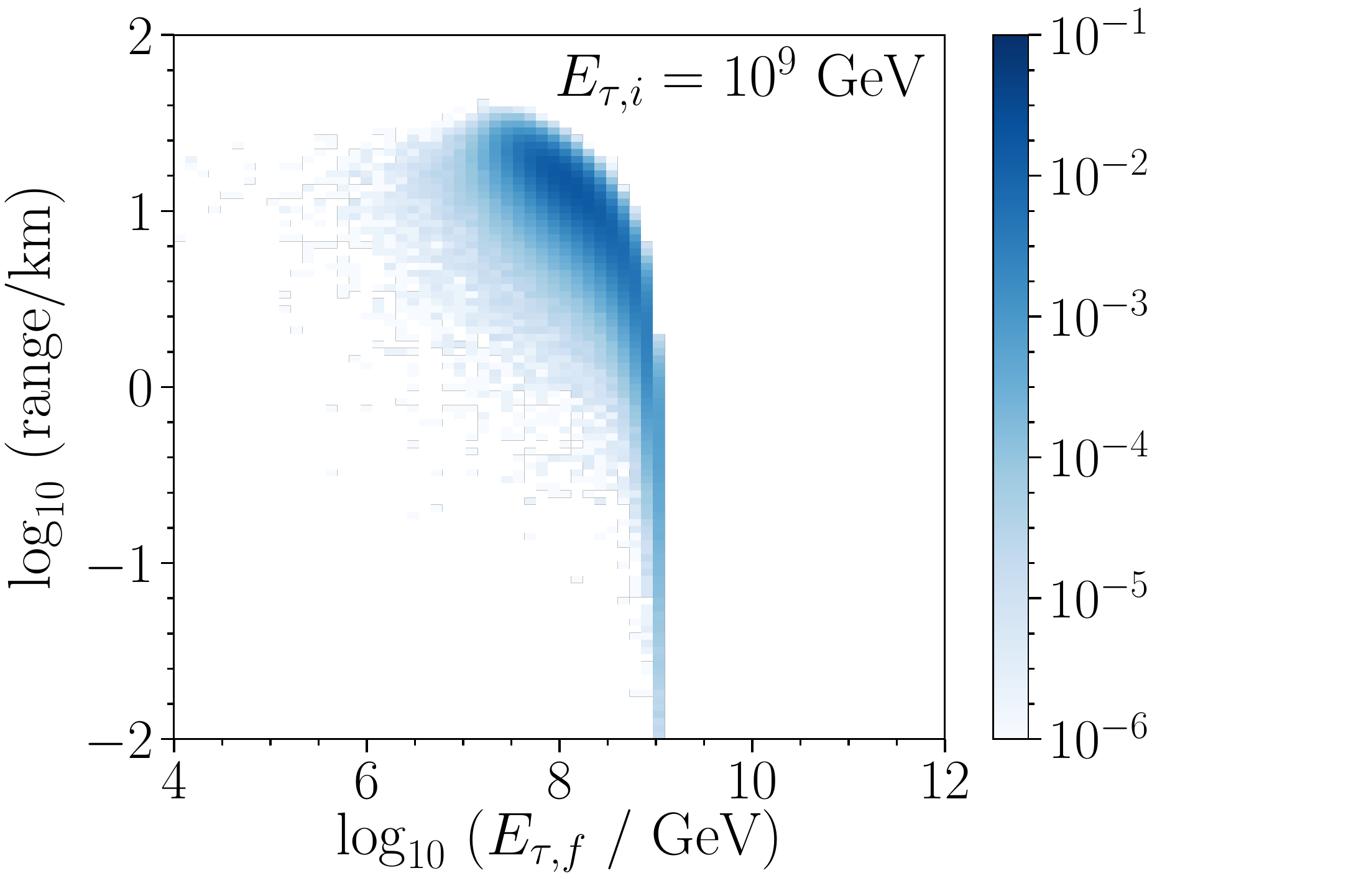}}\\
   \subfloat[][]{\includegraphics[width=.46\textwidth, trim={1.5cm 2cm 2cm 1cm}]{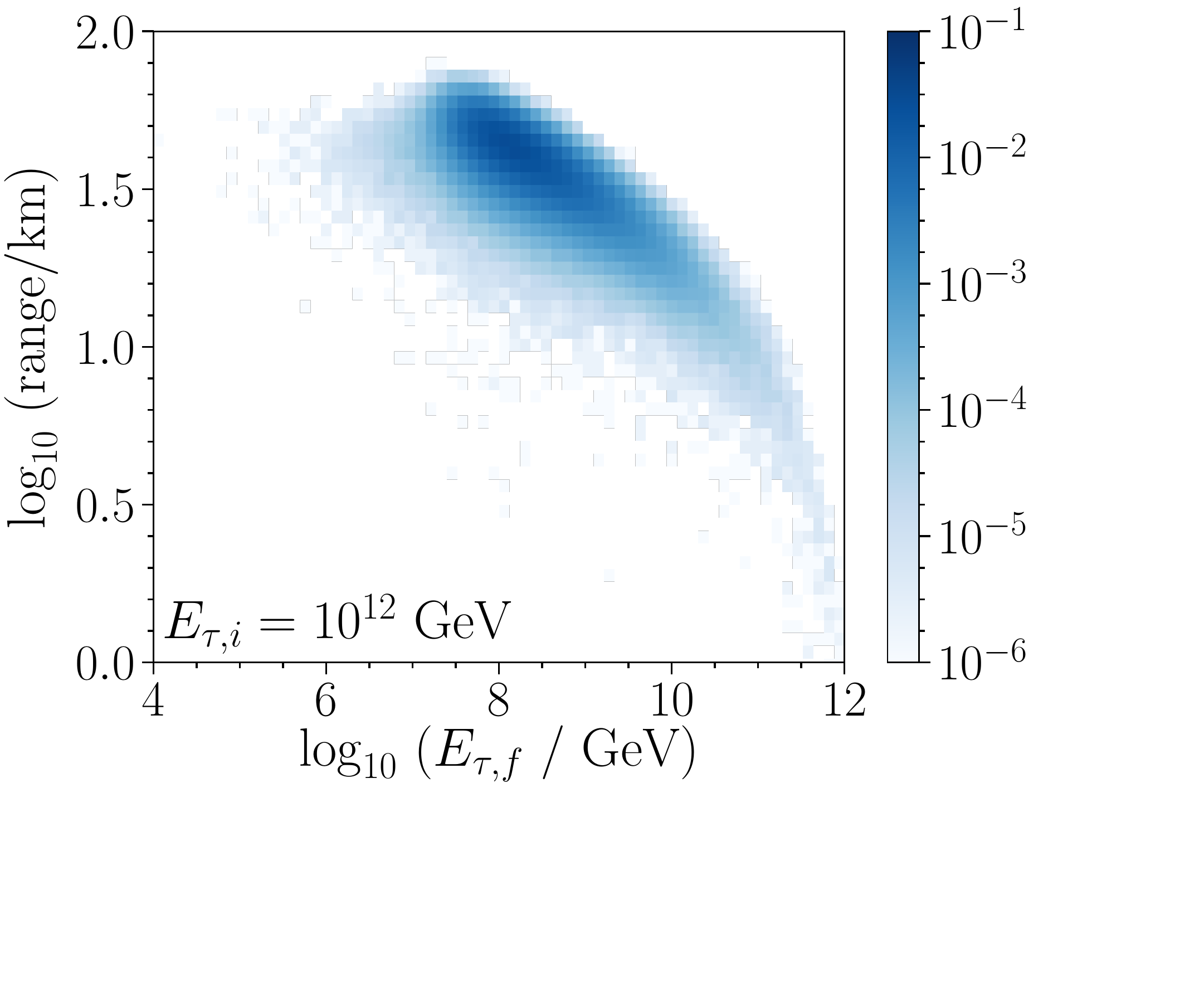}}\quad
   \subfloat[][]{\includegraphics[width=.46\textwidth, trim={1.5cm 2cm 2cm 1cm}]{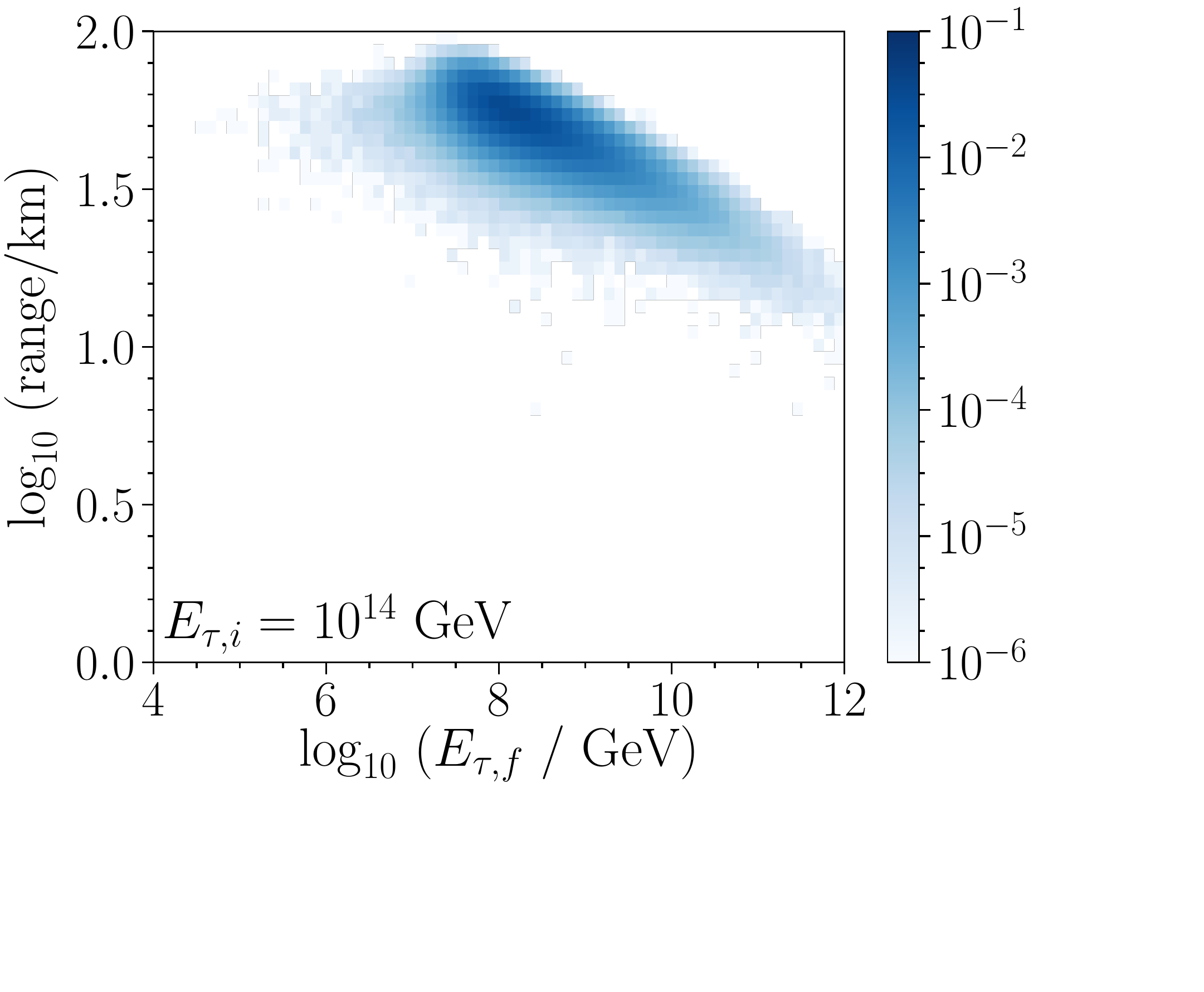}}
   \vspace{-1cm}
   \caption{Distribution of final tau energies and total distance traveled before decay for several initial tau energies. At 10 PeV (upper left) and below, taus lose little energy before decay, while at 1 EeV (upper right) taus reach the critical energy and losses become appreciable. In this regime, the median range increases linearly as the tau becomes more boosted. At 1 ZeV (bottom left) and above (bottom right), these distributions show asymptotic behavior, with taus decaying around 100 PeV and traveling, on average, tens of kilometers.}
   \label{fig:tau_losses}
\end{figure}

As discussed earlier, tau energy losses are negligible below 10 PeV and decay-on-the-spot is usually a good approximation.
Above the critical energy, taus lose energy through ionization, bremsstrahlung, pair production, and photo-nuclear interactions.
Ionization grows as $\ln(E_{\tau})$ and its contribution is minimal at the highest energies. Bremsstrahlung and pair-production have virtually no energy dependence above 1 PeV for taus and are sub-dominant, but are included in our treatment nonetheless. The photo-nuclear cross section grows with energy, and dominates the losses for taus above $\sim$1 EeV~\cite{Jeong:2017mzv}. This cross section depends on the nucleon structure function, and thus has the same source of uncertainty as the neutrino-nucleon cross section. For consistency, we use the same model of the nucleon structure function implemented for the neutrino-nucleon cross section. We incorporate it by modifying the publicly available Muon Monte Carlo (MMC) tool \cite{Chirkin:2004hz}, which we use to propagate taus. Fig.~\ref{fig:tau_losses} shows distributions of final tau energies and total distance traveled before decay for several initial tau energies.

\subsection{\texttt{TauRunner}\label{sec:tau_runner}}

\begin{figure}[htb!]
    \centering
    \includegraphics[width=\textwidth]{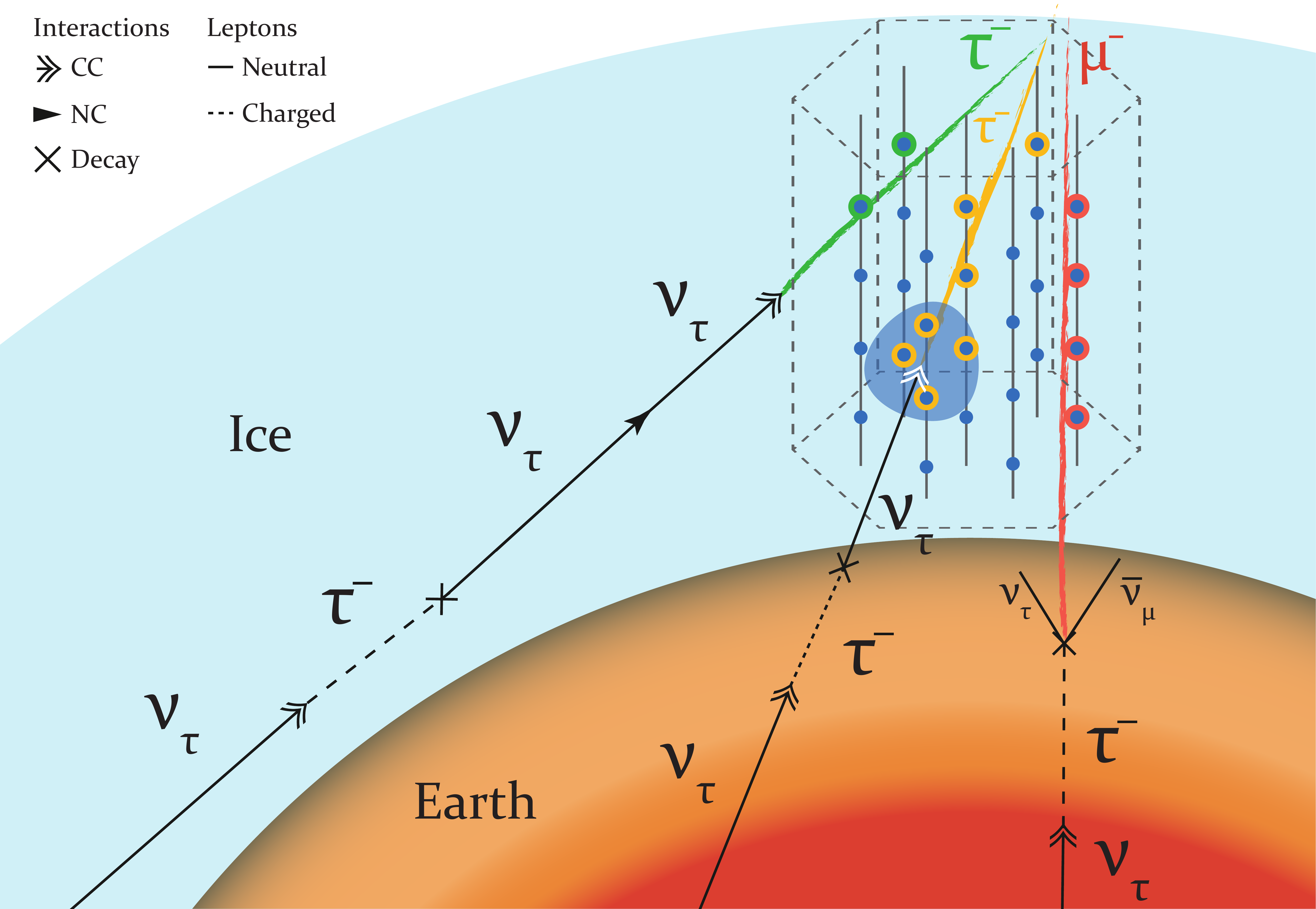}
    \caption{Schematic of lepton propagation through the Earth followed by a measurement with the IceCube detector. There are three possible signatures from EeV tau-neutrino secondaries, described here from left to right. Left: A through-going tau track, which is possible for taus at or above 10 PeV. Center: The interaction vertex is contained in the fiducial volume of the detector in this case, producing a cascade from the charged-current interaction, along with an outgoing tau track. Right: The tau decays before reaching the detector, producing a muon in ${\sim}18$\% of the cases, which subsequently enters the detector. For clarity, not all particles involved in the interaction are shown. An additional contribution included in the results but not shown here is a NC interaction inside the detection volume.}
    \label{fig:tau_runner_schematic}
\end{figure}


\texttt{TauRunner} is a Python package introduced for this analysis that propagates taus and neutrinos through a given medium, and is available at~\cite{TauRunner}. It begins by calculating the neutrino mean-free-path according to the total cross section and medium properties, followed by a random sampling to obtain the free-streaming distance. We use the Preliminary Reference Earth Model (PREM) \cite{Dziewonski:1981xy} for the density profile of the Earth, and compute the target number density using the isoscalar approximation. At the point of interaction, the specific process (NC or CC) is chosen via the accept or reject procedure. If the neutrino experiences a NC interaction, its energy loss is sampled from the differential cross section, and a new free-streaming distance is sampled. For CC interactions, a tau is created with an energy sampled from the corresponding differential cross sections. Tau energy losses, which include stochastic processes, are then calculated through a modified version of MMC. The tau final energy and distance traveled before decay are recorded. The tau-decay distribution for different modes has been parameterized in \cite{Dutta:2000jv}, from which we sample the energy of the daughter tau neutrino. The neutrino distributions described in \cite{Dutta:2000jv} depend on the polarization of the decaying tau. Taus produced in CC neutrino interactions are highly polarized~\cite{Hagiwara:2003di}. However, above 1 EeV, they undergo several interactions before decaying. As discussed earlier, the dominant interactions allowed before decaying are pair production~\cite{Koehne:2013gpa} and photo-hadronic interactions~\cite{Bugaev:2002gy}. These interactions are implemented in MMC~\cite{Chirkin:2004hz} by calculating the total cross section to all possible final states, which include those that change the tau polarization. This allows for the loss of the tau polarization after multiple scatterings. In order to take this into account, we take the simplifying assumption of considering taus produced above 1 EeV to be unpolarized at the point of decay. Below that energy, we average the negative and positive tau polarization distributions to account for neutrino and anti-neutrino propagation, respectively. From the tau-decay, only the leading tau neutrino is tracked and the process repeats. Propagation continues until the leading particle emerges, at which point the particle identity and final energy are recorded, along with a detailed history of undergone losses and interactions. A schematic showing the relevant outcomes of this process is shown in Fig.~\ref{fig:tau_runner_schematic}.

\section{Expected secondary neutrino distributions\label{sec:secondary_flux}}

\begin{figure}[ht!]
    \centering
    \includegraphics[width=0.95\linewidth]{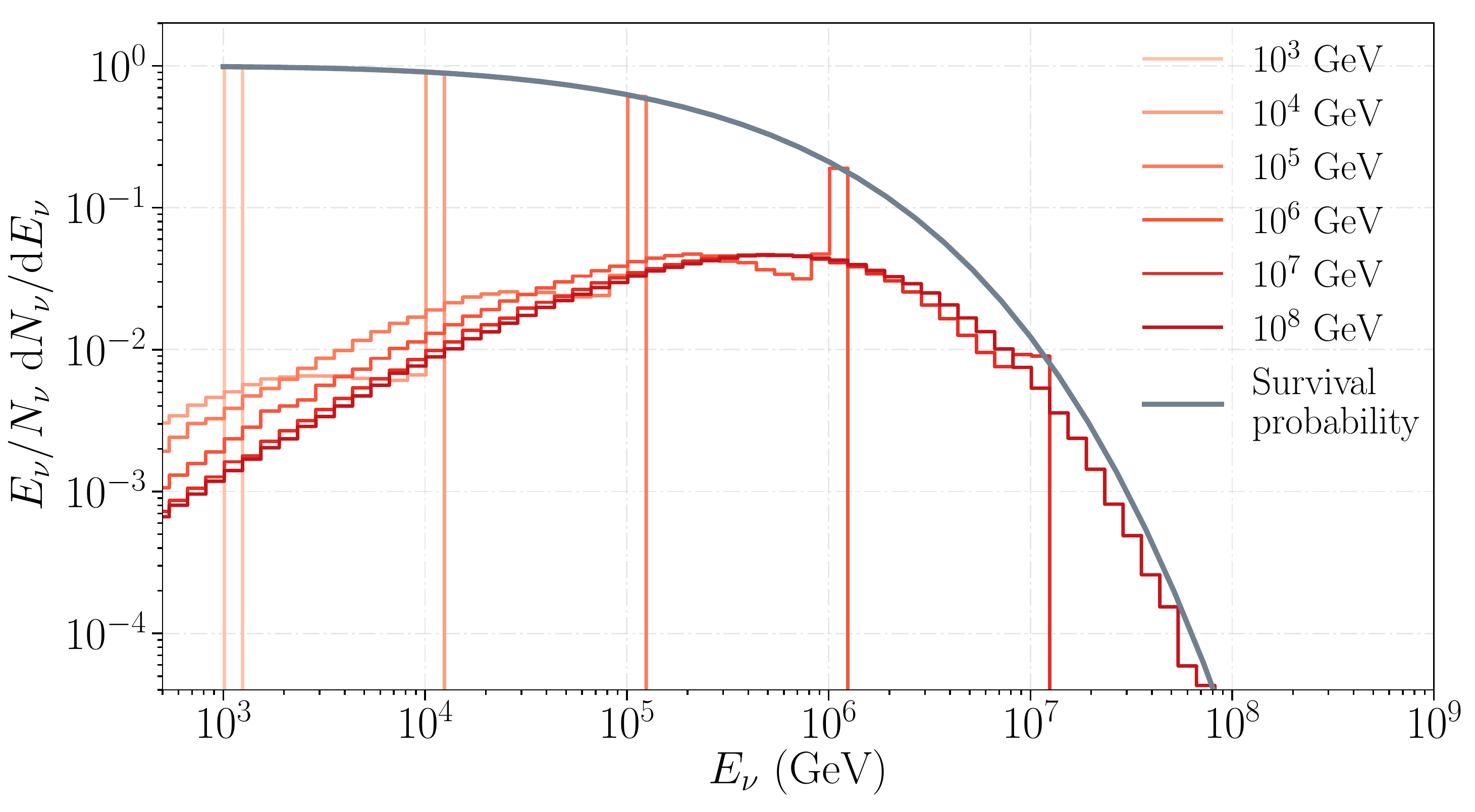}
    \caption{Mono-energetic tau neutrinos are injected at 30 degrees below the horizon, at a set of initial energies (specified in the legend), and propagated through Earth to calculate the resulting spectrum at emergence from the Earth. The rightmost bin in each spectrum represents the fraction of neutrinos that did not interact, while the secondary energy spectrum is represented by the curve to the left of each spike. The gray line shows the survival probability of the primary flux calculated for the same distance and column depth.}
    \label{fig:energy_dependence}
\end{figure}
\begin{figure}[ht!]
    \centering
    \includegraphics[width=\linewidth]{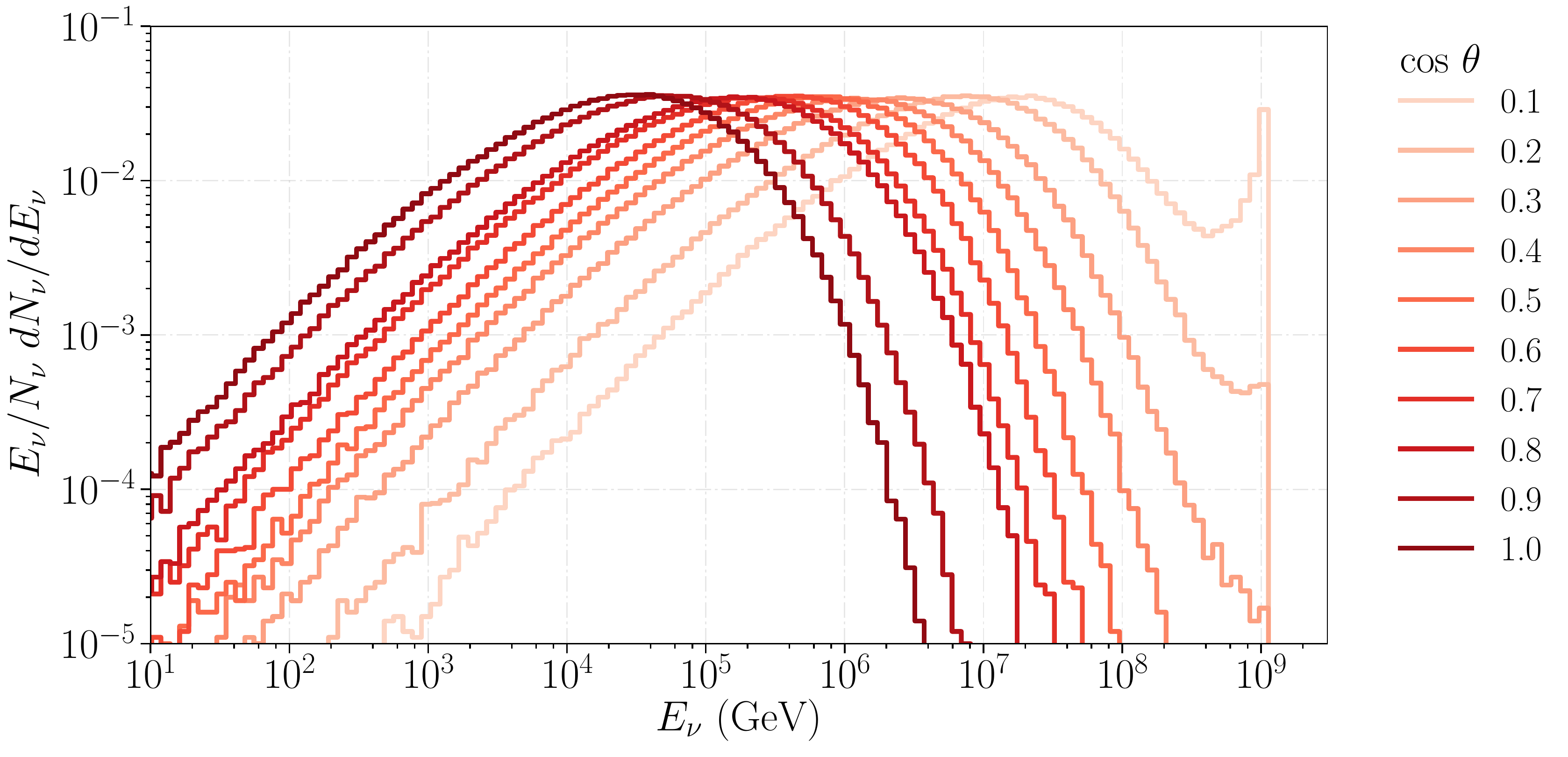}
    \caption{Tau neutrinos with an initial energy of 1 EeV with different incident angles are propagated through the Earth resulting in the cascaded tau neutrino spectra shown in red. At steep incident angles, exiting tau neutrino energies are centered around 100 TeV, while for shallower incidence, this peak is much higher in energy. The emergence angle $\theta$ is with respect to nadir.}
    \label{fig:zenith_dependence}
\end{figure}

We calculate the tau and neutrino energy distributions after traversing the Earth. We choose one energy value per decade from 100 GeV to 1 ZeV, and test a range of incident angles. Energy distributions from 1 TeV to 100 PeV are shown in Fig.~\ref{fig:energy_dependence}, and an angular distribution for 1 EeV neutrinos is shown in Fig.~\ref{fig:zenith_dependence}. We find that for angles greater than 20 degrees below the horizon and energies above ${\sim}1$ EeV, the secondary neutrino spectra are nearly identical. The reason for this primary energy degeneracy stems from the tau losses. As described in Sec.~\ref{sec:leptons_ehe}, the dominant energy losses grow with energy, which effectively means the tau loses more energy per column density traveled. This results in a flattening of the tau range, corresponding to the asymptotic behavior in Fig.~\ref{fig:tau_losses}. This effect, coupled with the short tau lifetime, causes the tau to travel roughly the same distance and decay around the same energy (10-100 PeV) regardless of its initial energy. We note this is counter-intuitive since one would expect (incorrectly) that a higher-energy incoming neutrino creates a higher-energy tau in a CC interaction, which would result in emerging neutrinos at higher energies. 

Therefore, the only differences in the secondary neutrino distributions are due to the variation of the first interaction length of the initial tau neutrino. For large enough column depths, this difference is negligible. For Earth-skimming neutrinos, however, the width of the distribution of the first interaction point is comparable to the corresponding column depth. An extended discussion of Earth-skimming neutrinos and their interactions can be found in \cite{Jeong:2017mzv, Dutta:2002zc, Venters:2019xwi, Reno:2019jtr, Dutta:2000jv, Dutta:2005yt}. 

Fig.~\ref{fig:energy_dependence} shows the secondary neutrino energy distributions after propagation through the Earth for a fixed angle of 30 degrees below the horizon. The gray line is the survival probability given by an exponential whose exponent is the ratio of the propagated distance to the neutrino mean interaction length. Thus, the rightmost bins in the distributions of Fig.~\ref{fig:energy_dependence} indicate the fraction of surviving primary neutrinos and, as expected, match the survival probability.

The most relevant feature of EeV tau neutrinos traversing the Earth are the energies with which they emerge. Fig.~\ref{fig:zenith_dependence} shows the distribution of outgoing events for an injected flux at 1 EeV, for several incident angles. Near the horizon, one can see the motivation for Earth-skimming detectors. These detectors are most sensitive to neutrinos that undergo a single CC interaction, which is inferred through the detection of the subsequent tau decay shower in the atmosphere. However, at steeper angles, it becomes less likely for a tau to exit the Earth. It's much more likely that the tau will instead decay in the Earth, producing a tau neutrino with energy between 100 TeV and 10 PeV. This is the regime where cubic-kilometer neutrino detectors such as IceCube effectively operate. Thus, there is an opportunity to study cosmogenic fluxes via the detection of cascaded daughter particles. This will be discussed in further detail in Sec.~\ref{sec:GZK}.

\label{sec:secondaries}


\section{Applications and Implications\label{sec:applications}}

\noindent IceCube has measured the diffuse neutrino flux at energies extending to $\sim$10 PeV, and has placed upper limits at higher energies \cite{Kopper:2017zzm, Aartsen:2016xlq, Aartsen:2015knd, Aartsen:2014gkd}. Although IceCube is sensitive to EHE fluxes directly, and has set limits in that range, these searches are limited to a small region near the horizon since most of the primary flux is lost beyond that. As was discussed previously, radio detectors were developed to look for downgoing as well as near-horizon EeV neutrinos as they skim the Earth, yet no claim of cosmogenic neutrino detection has been made. Two exceptions are the anomalous ANITA events, which were detected at much steeper angles than would be expected from an isotropic neutrino flux. We discuss one of these events here in the context of its predicted PeV counterparts at IceCube. We then study the PeV counterparts of neutrinos from a diffuse cosmogenic flux and show the expected signal in ten years of IceCube data.


\subsection{ANITA and its anomalous events\label{sec:ANITA}}

The ANITA collaboration has reported the detection of two events whose signatures are consistent with upgoing air showers produced by a tau \cite{Gorham:2016zah,Gorham:2018ydl}. This interpretation requires the decay of a tau (from a tau neutrino CC interaction) to occur in the atmosphere producing an extensive air shower (EAS). This is distinguishable from a reflected EAS initiated by a cosmic-ray, in which the radio signal acquires a phase reversal from reflection off of the Antarctic ice, while an upgoing EAS does not display such a phase reversal. However, this interpretation is problematic as tau neutrinos with energies to which ANITA is sensitive are not likely to travel through the large Earth column depths required for these events. While it has been noted that these events are unlikely to be caused by an isotropic neutrino flux \cite{Romero-Wolf:2018zxt, Fox:2018syq, Chipman:2019vjm}, discrete-source emission could evade these constraints. Beyond the Standard Model (BSM) explanations have also been proposed. This includes axion-photon conversion \cite{Esteban:2019hcm}, sterile neutrinos \cite{Chipman:2019vjm,Huang:2018als, Anchordoqui:2018ucj, Cherry:2018rxj}, and heavy SUSY partners or Dark Matter particle decays \cite{Connolly:2018ewv, Collins:2018jpg, Anchordoqui:2018ssd, Heurtier:2019git, Hooper:2019ytr, Cline:2019snp, Heurtier:2019rkz, Borah:2019ciw}. Here, we examine the discrete-source emission hypothesis and show that any detection of EeV neutrinos from steep incident angles at ANITA can be ruled out by the non-observation of TeV - PeV neutrinos with other neutrino telescopes, such as IceCube. 

\begin{figure}[ht!]
    \centering
    \includegraphics[width=0.55\linewidth]{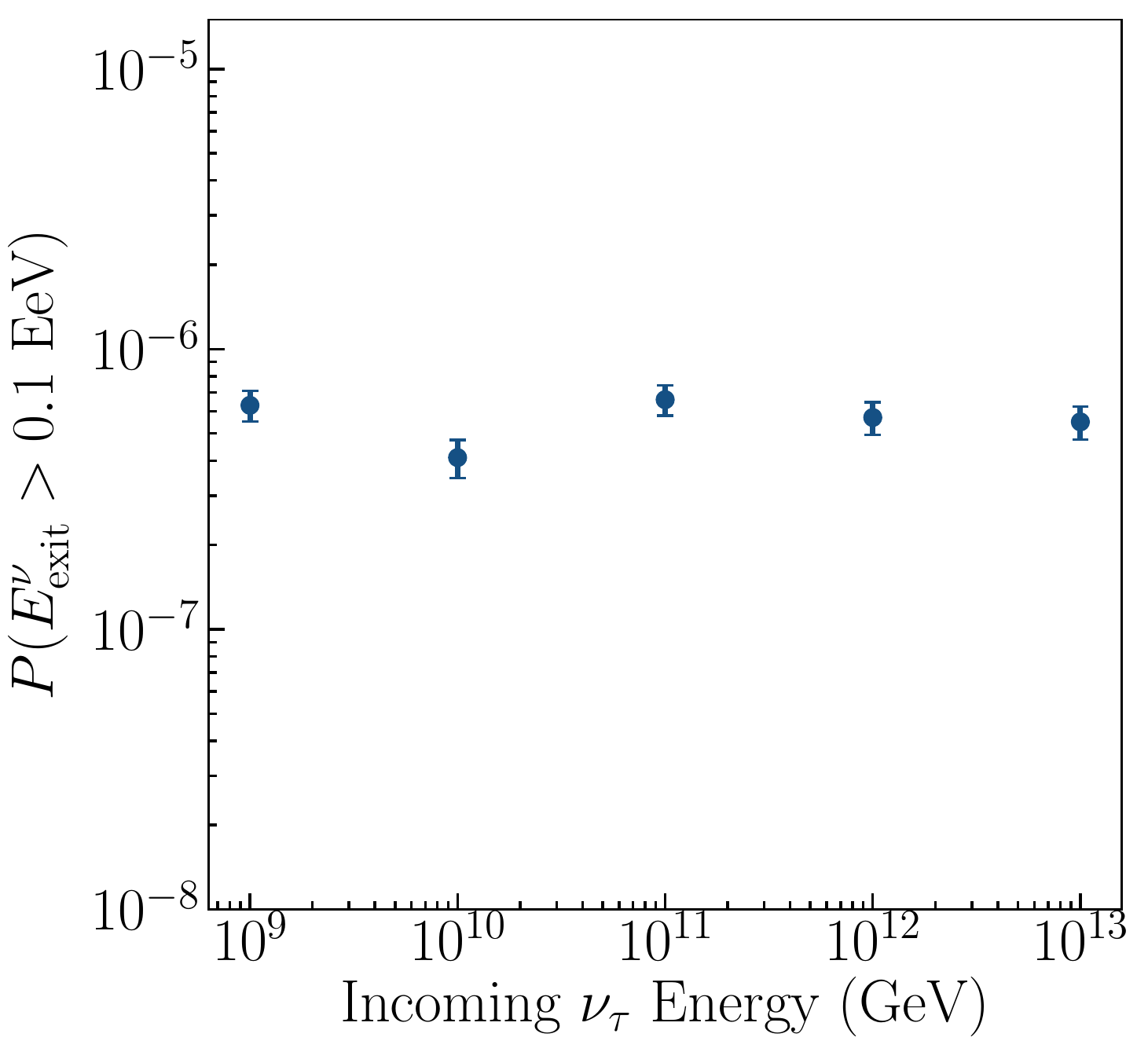}
    \caption{Probability for 
    a tau neutrino to exit Earth with a minimum energy of 0.1 EeV (approximate ANITA threshold), after Earth propagation (based on inferred chord length for AAE141220), assuming $\nu_{\tau}$ incidence at a particular initial energy. Errors are statistical only.}
    \label{fig:tau_prob}
\end{figure}

The number of events detected by ANITA due to tau showers in the atmosphere from a primary neutrino flux, $\Phi \left( E _ { \nu } \right)$, is given by

\begin{equation} \label{eq:anita_exp}
    \mathcal{N} _ { \nu } = \int d E_{ \nu } d E^{\prime} _ { \nu } ~ \Phi \left( E _ { \nu } \right) \frac { d N_{\nu} } { d E' _ { \nu } } \left(E^{\prime} _ { \nu } ; E_ { \nu } \right) \xi _ { a c c } \left( E' _ { \nu } \right) \Delta T \; ,
\end{equation}
where $E_{ \nu }$ and $E^{\prime} _ { \nu }$ are the primary and secondary neutrino energy, respectively. $ d N_{\nu} \left( E' _ { \nu };  E _ { \nu }  \right) /  d E' _ { \nu } $ is the energy distribution of secondary tau neutrinos near the ice surface, $\Delta T$ is the duration of observation, and $\xi _ { a c c } \left( E ' _ { \nu } \right)$ is the ANITA acceptance \cite{Romero-Wolf:2018zxt} in units of cm$^2$sr. The acceptance incorporates the probability of neutrinos interacting in the ice, as well as the probability of a tau decay shower occurring in the atmosphere. Given that the reported acceptance in  \cite{Romero-Wolf:2018zxt} includes neutrino propagation through the Earth, we set the acceptance at all angles to be that near the horizon to remove the Earth absorption effects, which we account for separately with \texttt{TauRunner}. For the incoming flux, we take the minimalistic assumption of a delta function in energy, $\Phi \left( E _ { \nu } \right) = \frac{dN}{dA dt dE_{\nu}} = \Phi _ { 0 } \delta \left( E _ { \nu } - E _ { 0 } \right)$, where $\Phi_{0}$ is the normalization with units $\rm{cm}^{-2} \rm{s}^{-1}$. Probabilities of tau neutrinos 
exiting the Earth with energies greater than 0.1 EeV are shown in Fig.~\ref{fig:tau_prob}, for the chord lengths corresponding to AAE141220. For both taus and tau neutrinos, the probability of exiting the Earth with an energy larger than 0.1 EeV seems to be fairly independent of energy, for initial tau neutrino energies above 1 EeV. Therefore, in what follows, we choose $E_0 = 1$ EeV as the primary energy. Details of this primary energy degeneracy are discussed in more detail in Sec.~\ref{sec:secondary_flux}.

\begin{figure}[]
    \centering
    \includegraphics[width=0.75\linewidth]{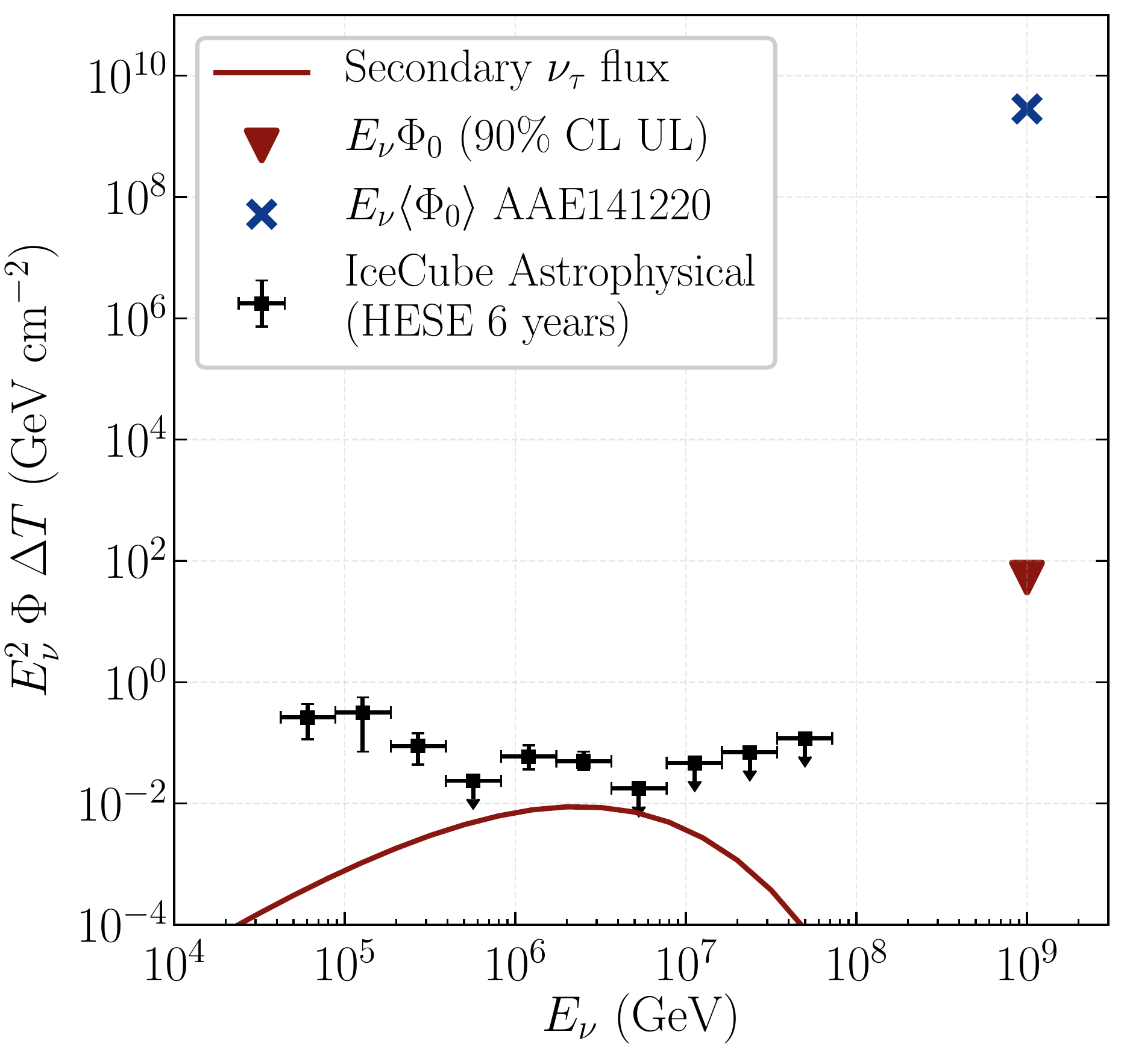}
    \caption{Maximum allowed flux of EeV neutrinos (maroon arrow), given an injected mono-energetic neutrino flux at or above the detected ANITA event (AAE141220) energy. The normalization of the secondary flux is set to the maximum that does not exceed IceCube's diffuse astrophysical flux (black bins). The flux needed to produce one event in the third flight of ANITA (blue marker) exceeds the upper limit by many orders of magnitude. We use the published spectrum based on six years of high energy starting events.}
    \label{fig:ANITA_UL}
\end{figure}

As was discussed above, this primary flux of EeV neutrinos is guaranteed to be associated with a secondary flux of TeV to PeV neutrinos. Such a large rate of TeV muons crossing the IceCube volume simultaneously would deposit a large amount of charge. Large-charge events are promptly reported by IceCube via the EHE and HESE streams\footnote{\url{https://gcn.gsfc.nasa.gov/amon.html}}. For example, the EHE stream requires three thousand photo-electrons and thirty channels to trigger an alert. A $\sim$1 PeV muon typically deposits $\sim$200 TeV, which on average results in over four thousand photo-electrons in more than 40 channels, when crossing the full detector \cite{jvs}. In fact, the largest energy deposition reported in these streams corresponds to 5 PeV deposited energy\footnote{\url{https://gcn.gsfc.nasa.gov/gcn3/24028.gcn3}}, but it's for a down-going event; horizontal and upward going events have not had multi-PeV announcements in these streams. Thus, we conclude that IceCube has not observed catastrophic events that would be produced by bundles of TeV neutrino-induced muons. In what follows, we then take the conservative assumption that a single muon makes it through. Such events have been observed and we can compare this expected yield to IceCube's measurement of the high-energy events.
We find the maximum allowed normalization of the incident flux by comparing the secondary neutrino distribution with the measured IceCube astrophysical flux from the High-Energy Starting Event selection (HESE) \cite{Kopper:2017zzm}. Results for AAE141220 are shown in Fig. \ref{fig:ANITA_UL}. The unfolded HESE spectrum is folded back to the detector using \texttt{TauRunner} as discussed in Sec.~\ref{sec:tau_runner}.

The 90\% C.L upper limit on the EeV primary flux normalization is set by comparing both secondary distributions and requiring that the secondaries produced by the primary EeV flux do not exceed those of HESE at 90\% C.L. Given that the time profile of the intrinsic flux is unknown, we place limits on the time-integrated flux. We take the duration to be 22 days ($\Delta T$ in Eq.~\eqref{eq:anita_exp}), corresponding to the entire ANITA-III flight. We find the maximum allowed time-integrated flux to be $E^2 \Phi \Delta T \simeq 10^2$ GeV cm$^{-2}$. Using the maximum allowed time-integrated flux, we calculate the expected number of events at ANITA. This yields a maximum expected number of neutrinos of less than $ \mathcal{O}\left(10^{-7}\right)$ in 22 days. 
This is illustrated in Fig.~\ref{fig:ANITA_UL} where we show the flux required to produce one event at ANITA as a reference. It is therefore highly unlikely for the reported event to be caused by a high-energy tau neutrino. 

In this analysis, we integrate the IceCube measurement of the astrophysical flux over 22 days. However, as this measurement was made over six years, we are working under the assumption that the astrophysical flux has no large dependence on time. It is worth noting, however, that short timescale transients are allowed to overproduce the measured astrophysical flux, so long as they do not overproduce the astrophysical flux integrated over the duration of the measurement. For this reason, a dedicated analysis by IceCube searching for short timescale emission around the time of the ANITA event ought to be performed.

\subsection{Cosmogenic Flux}
\label{sec:GZK}
At energies beyond the so-called Greisen-Zatsepin-Kuzmin (GZK) cutoff ($E \geq 40$ EeV), proton interactions with the CMB restrict the mean-free path of cosmic-ray nuclei primaries to less than a few hundred Mpc from sites of cosmic acceleration. A suppression compatible with the GZK cutoff has indeed been observed in cosmic-ray experiments \cite{Abbasi:2007sv, Sokolsky:2008zz, Abraham:2008ru, AbuZayyad:2012ru}. The subsequent decay of the mesons from these interactions leads to an observable, yet currently undetected, flux of cosmogenic neutrinos. 


Although the cosmogenic flux should be isotropic at Earth, searches for this flux have been limited to either half of the sky (downgoing) or small solid angles, specifically looking for Earth-skimming neutrinos, where the probability of detecting a tau in the atmosphere after a single neutrino interaction in the Earth is optimized \cite{Abbasi:2010ak, Barwick:2014pca, Aab:2015kma, Aartsen:2016ngq, Allison:2015eky,Allison:2018cxu,Aartsen:2018vtx,Aab:2019auo, Kotera:2010yn}. Here, we show how using the secondary flux will extend this search to the entire sky. Specifically, we look for neutrinos after several interactions in the Earth, which we have shown emerge at $\mathcal{O}$(PeV) energies. We also show that the rate from Earth-traversing neutrinos is not negligible. 

\begin{figure}
    \centering
    \includegraphics[width=0.85\textwidth]{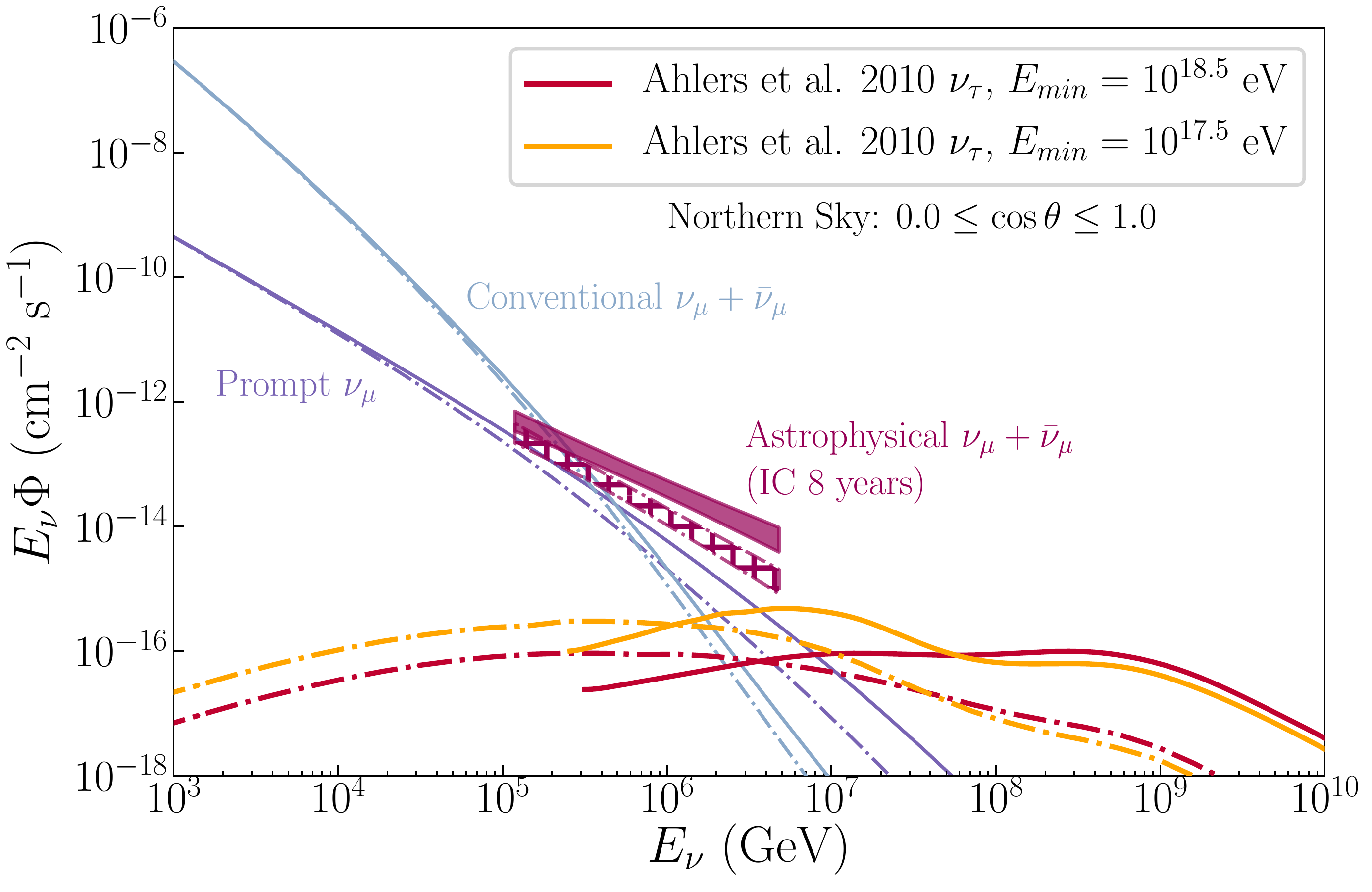}
    \caption{Per flavor neutrino fluxes from 1 TeV to 10 EeV, integrated over the Northern Sky. Primary fluxes are shown in solid lines, and fluxes present at the IceCube detector are shown in dashed-dotted lines and hatches. $\nu_{\tau}$ components for various models of the cosmogenic flux are shown in red and orange \cite{Ahlers:2010fw}. These spectra are compared to models of both the conventional \cite{Honda:2006qj} and prompt \cite{Enberg:2008te} components of the atmospheric flux 
    as well as measurements of the diffuse astrophysical flux \cite{Haack:2017dxi}.  Secondary $\nu_{\tau}$ spectra peak at PeV energies, a region of parameter space optimal for neutrino telescopes such as IceCube.}
    \label{fig:gzk_and_friends_wholesky}
\end{figure}

Fig.~\ref{fig:gzk_and_friends_wholesky} displays the secondary tau neutrino flux of cosmogenic neutrinos compared to atmospheric and diffuse astrophysical per-flavor neutrino fluxes in the Northern sky. For a cosmogenic flux, we choose a model produced from a fit to HiRes data \cite{Ahlers:2010fw}. The conventional component in Fig.~\ref{fig:gzk_and_friends_wholesky} shows the $\nu_{\mu}$ flux produced in cosmic-ray showers in the atmosphere, using the model in \cite{Honda:2006qj}. The prompt component is the expected muon neutrino flux arising from atmospheric charm production in cosmic-ray showers; we use the model in \cite{Enberg:2008te}. Although there is a predicted $\nu_{\tau}$ component of this prompt flux, predominantly from D-meson decays, the level of this flux is much smaller compared to the prompt $\nu_{e,\mu}$ components. The astrophysical muon-neutrino flux we use is based on eight years of Northern sky muon track data from IceCube \cite{Haack:2017dxi}. All of these primary fluxes are propagated to the detector. The fluxes arriving at the detector are then compared to the secondary flux from cosmogenic neutrinos. The spectrum of the secondary cosmogenic flux is much harder and strongly dependent on declination, providing additional handles to distinguish cosmogenic secondaries from other astrophysical or atmospheric events. 

\begin{figure}[!ht]
    \centering
    \includegraphics[width=0.9\textwidth,trim={0cm 0cm 0cm 0cm}]{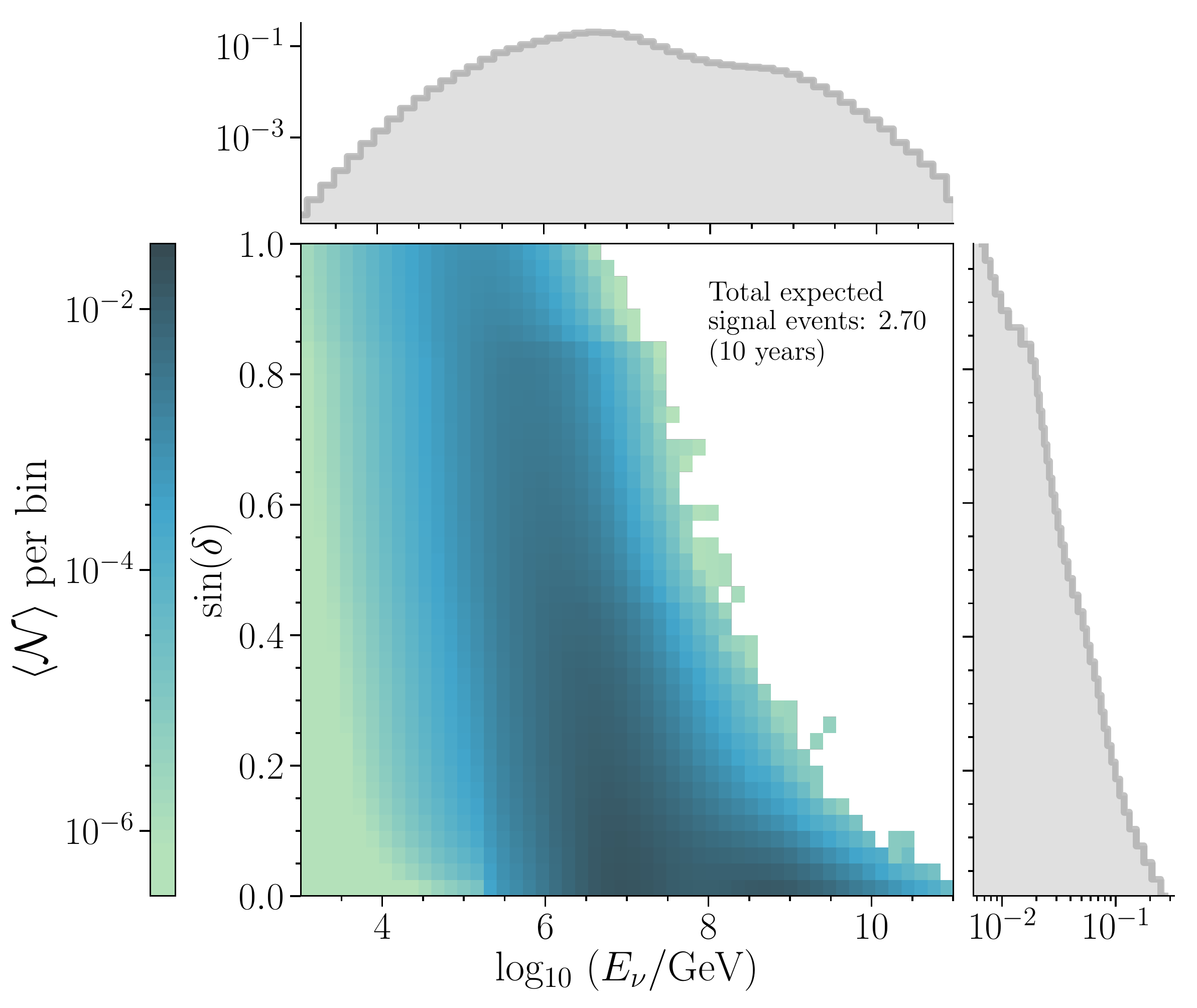}
    \caption{Expected signal of cosmogenic neutrinos at IceCube, assuming the model from \cite{Ahlers:2010fw} and assuming a cosmic-ray composition that is dominated by protons above energies of $10^{17.5}$ eV, after 10 years of data collection. The Earth-skimming contribution represents only about one third of the total expectation, and the majority of events are expected to peak around 10 PeV in true neutrino energy.
    }
    \label{fig:GZK_signal}
\end{figure}

To further highlight the expected signal shape, we show the resulting expected signal distribution of this benchmark model in Fig.~\ref{fig:GZK_signal}. The number of expected signal events at IceCube is calculated by propagating a $\nu_{\tau}$ flux isotropically over the Northern hemisphere from incidence on the Earth to a few kilometers away from IceCube. The number of events expected at IceCube is then given by

\begin{equation}
    \mathcal{N}_{\nu}^{\rm{GZK}} ~ = \int dE^{\prime} d\Omega ~ \Phi_{\nu} (E^{\prime}_{\nu}) \Delta T \left[ \sigma_{\nu N}^{CC} (E_{\nu}^{\prime})  \cdot ~ \frac{\Gamma_{\tau \rightarrow \mu}}{\Gamma_{total}} \cdot N^{CC}_{N}(E_{\nu}^{\prime}) + \sigma_{\nu N}^{NC}(E_{\nu}^{\prime}) \cdot N^{NC}_{N} \right], 
\end{equation}
where $ \Phi_{\nu} (E'_{\nu})$ is the emerging flux near the detector, $\sigma_{\nu N}^{CC}$ and $\sigma_{\nu N}^{NC}$ is the neutrino-nucleon isoscalar cross section for charged- and neutral- current, respectively. $\Gamma_{\tau \rightarrow \mu} / \Gamma_{total}$ is the tau to muon branching fraction, and $N_{N}$ is the effective number of isoscalar targets. This number is fixed to be $N$ targets in 1 km$^3$ of ice for the NC channel, but has an energy dependence for the CC channel due to the extended muon range, and is given by,

\begin{equation}
    N^{CC}_{N} \left( E_{\nu} \right) = \int d\tilde{E_{\mu}}d\tilde{E}_{\tau}  \frac { d N_{\tau} } { d \tilde{E}_{ \tau} } \left(\tilde{E} _ { \tau } ; E^{\prime}_ { \nu } \right) \frac { d N_{\mu} } { d E_{ \mu} } \left(E_{\mu} ; \tilde{E} _ { \tau } \right) R_{\mu} \left(E_{\mu} \right) A^{geo} \frac{\rho^{ice}}{M_{iso}}.
    \label{eq:ncc}
\end{equation}

\noindent The first and second term in Eq.~\ref{eq:ncc} are the tau and muon energy distributions, respectively, $R_{\mu} \left(E_{\mu} \right)$ is the average muon range calculated with MMC,  $A^{geo}$ is the geometrical transverse area (1 km$^2$ in this case), $\rho^{ice}$ is the density of ice, and $M_{iso}$ is the isoscalar nucleon mass. Fig.~\ref{fig:GZK_signal} shows the expected number of events at IceCube binned in true neutrino energy and declination. We find that, assuming a proton-dominated UHECR flux with a minimum crossover energy of 10$^{17.5}$ eV (10$^{18.5}$ eV), IceCube should see 2.70 (1.25) upgoing neutrinos with a hard energy spectrum, peaking at 10 PeV, in ten years of data taking. These events are dominated by the CC channel, with only around 10\% of the signal coming from NC interactions in the fiducial volume of the detector. Of all of the events, we find that only $\sim$0.8 (0.5) would be Earth-skimming, where we have defined Earth-skimming to be up to 5 degrees below the horizon. Therefore, in total, we expect the rate from Earth-traversing neutrinos to be at least twice that from Earth-skimming neutrinos.

\begin{figure}
    \centering
   \subfloat[][]{\includegraphics[width=.46\textwidth, trim={1cm 1.25cm 1cm 1cm } ]{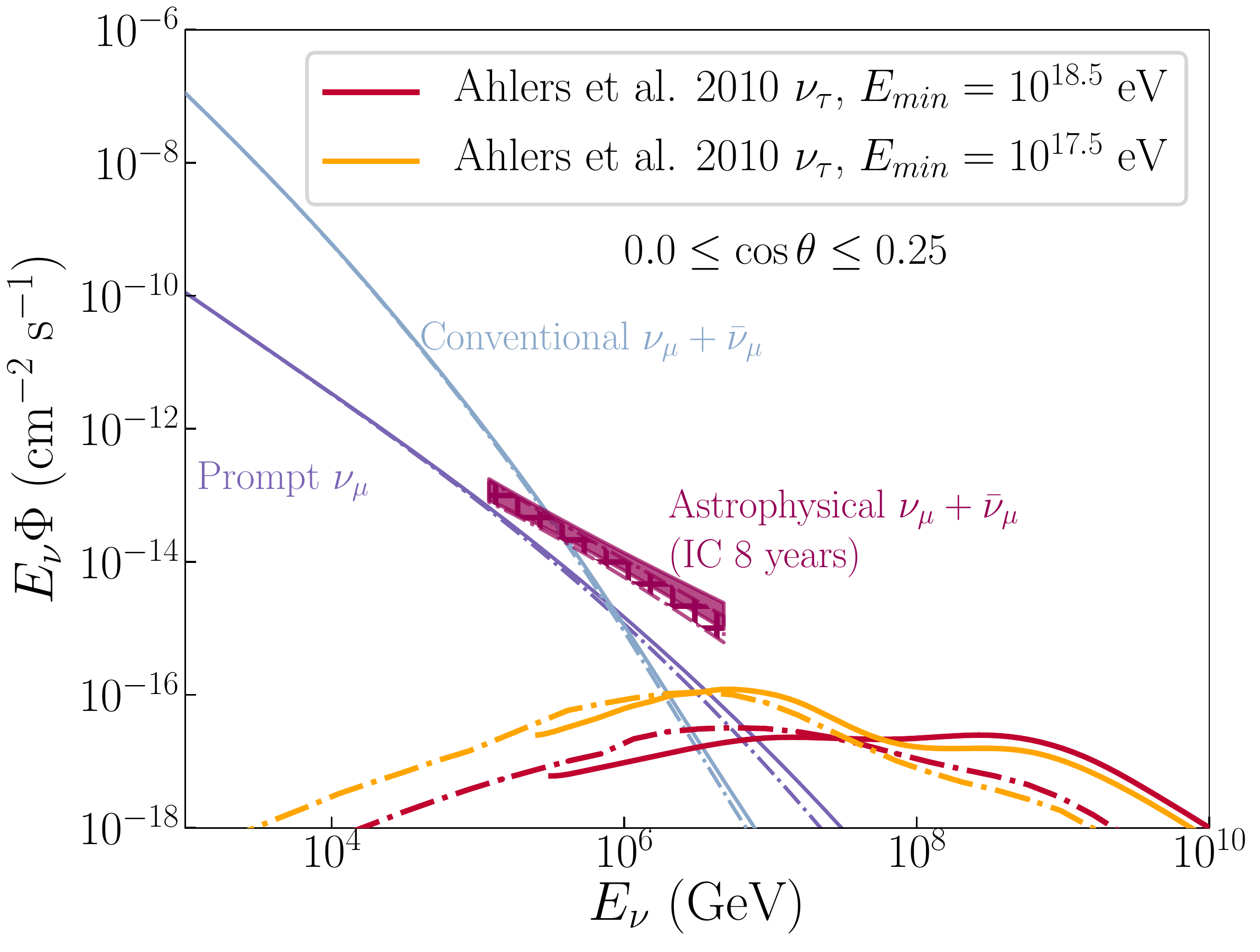}}\quad
   \subfloat[][]{\includegraphics[width=.46\textwidth, trim={1cm 1.25cm 1cm 1cm }]{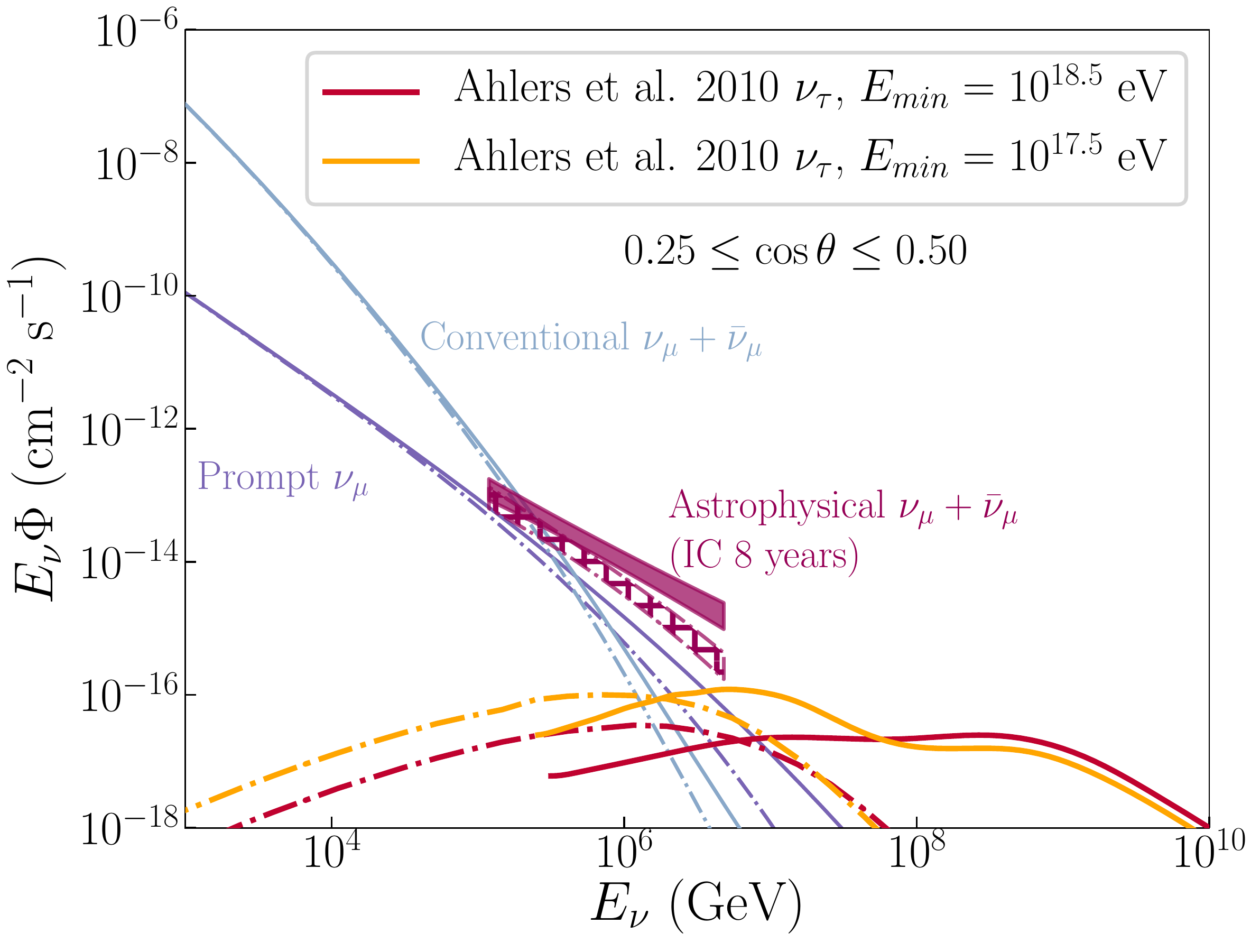}}\\
   \subfloat[][]{\includegraphics[width=.46\textwidth, trim={1cm 1cm 1cm 1.25cm }]{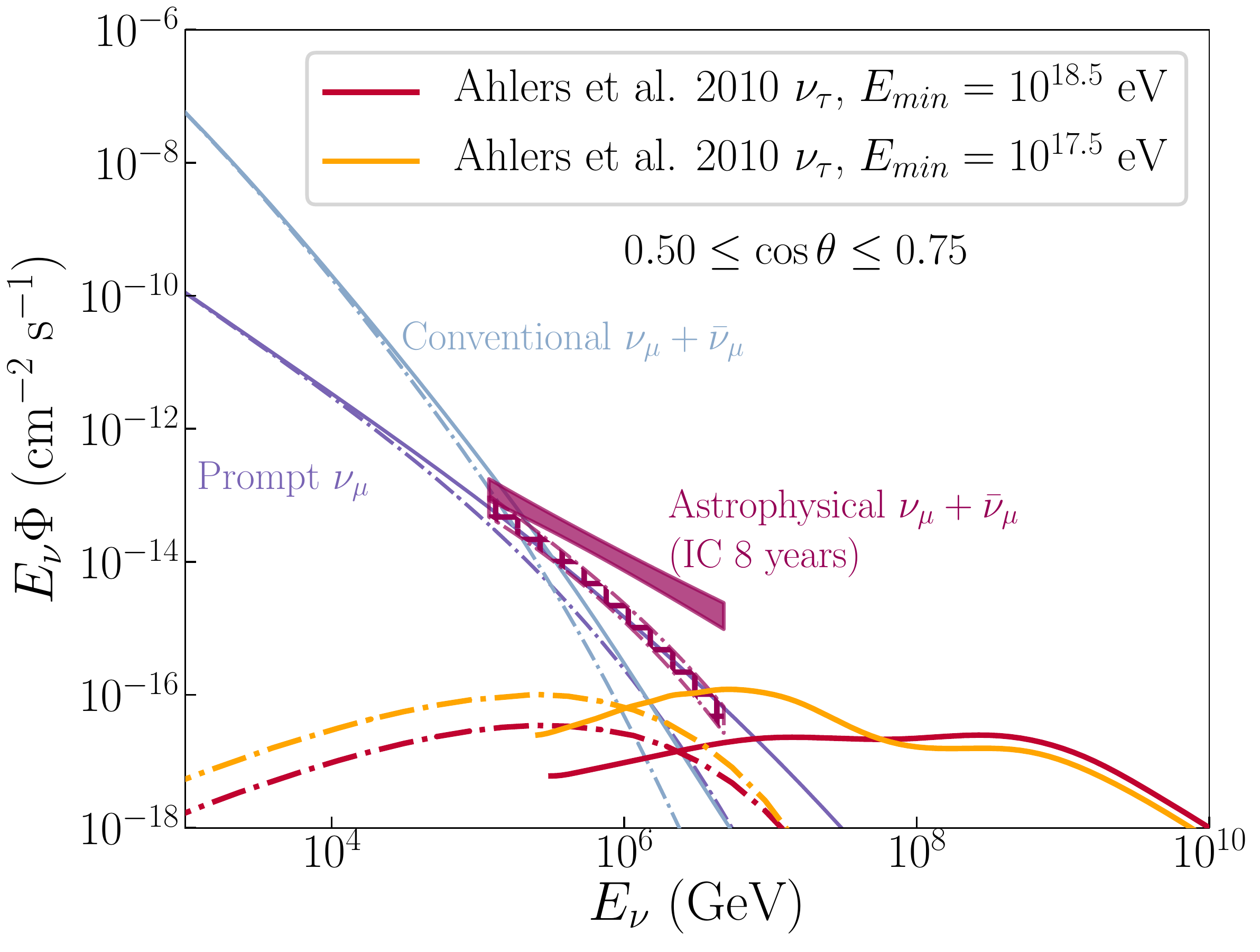}}\quad
   \subfloat[][]{\includegraphics[width=.46\textwidth, trim={1cm 1cm 1cm 1.25cm }]{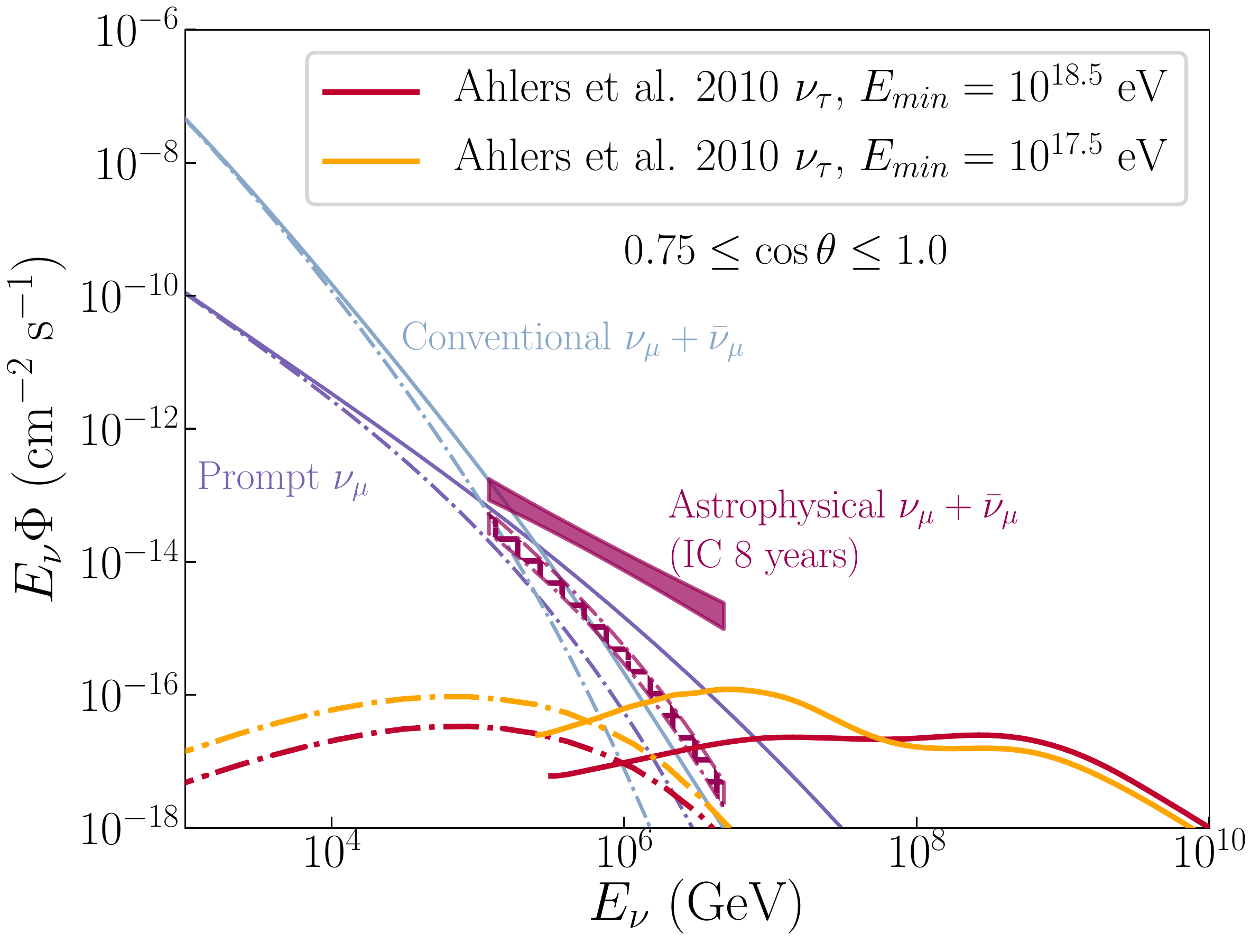}}\\
    \caption{Per flavor neutrino fluxes from 1 TeV to 10 EeV, integrated over various zenith bands in the Northern Sky. Solid lines are primary fluxes, while secondary fluxes are represented by dashed-dotted lines and hatches. The secondary cosmogenic tau neutrino spectrum is strongly dependent on the incoming zenith angle. For arrival directions towards Earth's core, it contributes equally to the astrophysical flux at IceCube above 2 PeV.}
    \label{fig:gzk_and_friends_bands}
\end{figure}

Fig.~\ref{fig:gzk_and_friends_bands} further demonstrates the declination dependence of this flux through comparison to atmospheric backgrounds, and shows that in certain zenith angle bands with large enough column depth through the Earth, the flux arriving at IceCube is higher than the atmospheric background at and above 2 PeV. It is important to note that testing several cosmogenic models showed that this technique yields more expected events when the primary fluxes peak at energies below 1 EeV. While cosmogenic neutrino fluxes predicted from heavy cosmic-ray nuclei primaries suppress the component at and above 1 EeV, they boost the expectation between 1 and 100 PeV. Therefore, this indirect detection method can prove essential if the cosmic-ray primary composition is determined to be dominated by heavy nuclei.

\section{Discussion and Conclusion}

In this work we have introduced a new Monte Carlo package, \texttt{TauRunner}, to propagate EHE neutrinos and taus, including updated cross section models and tau energy losses. We apply this calculation to two interesting cases. In the first case, we consider the anomalous ANITA events and find that the maximum allowed secondary neutrino flux constrained by IceCube measurements implies a primary flux that is inconsistent with a Standard Model neutrino explanation of AAE141220. We calculate that ANITA should see less than $\mathcal{O}\left(10^{-7}\right)$ events in the reported direction during the entire third flight, requiring a significant over-fluctuation to detect one event. As shown above, this conclusion is independent of the incident spectral shape and time profile. In the second case, we propagate cosmogenic neutrino fluxes through the Earth and find that the secondary flux of TeV-PeV neutrinos from cascaded cosmogenic fluxes is a non-negligible contribution to the total astrophysical flux at IceCube. Depending on the primary model used, this secondary flux can reach 20\% of the total flux at the detector above a PeV. We calculate the expected number of events from secondary cosmogenic neutrinos as well as their energy and zenith distributions. We find that the expected rate at IceCube from secondary neutrinos is twice the rate at the horizon, albeit at lower energies where the astrophysical background is higher. In the future, the larger effective area of IceCube Gen2 will allow a dedicated IceCube analysis to fit for this signal using its joint spectral and angular distribution and provide a complementary measurement to detectors optimized for the EeV scale.

\label{sec:conclusion}

\acknowledgments
We would like to thank Segev BenZvi, Dmitry Chirkin, Sam Fahey, Mary-Hall Reno, Subir Sarkar, and Abigail Vieregg for useful discussions. We thank Kareem Farrag for creating the schematic in Figure 3, and for useful discussions. CA is supported by NSF grant PHY-1801996. FH, RH, AK, AP, IS, and JV are supported by NSF under grants PLR-1600823 and PHY-1607644 and by the University of Wisconsin Research Council with funds granted by the Wisconsin Alumni Research Foundation.

\bibliographystyle{JHEP}
\bibliography{references}

\end{document}